# Uncertainty evaluation of length measurement by spectral domain interferometry with super-Gaussian windowing

**Dae Hee Kim[1,2], Yoonkwon On[1,2], Sunghoon Eom[1], Chu-Shik Kang[1,2], and Yoon-Soo Jang[1,2,3,*]**

[1] Length and Dimensional Metrology Group, Division of Physical Metrology, Korea Research Institute of Standards and Science (KRISS), 267 Gajeong-ro, Yuseong-gu, Daejeon, 34113, Republic of Korea
[2] Department of Precision Measurement, University of Science and Technology (UST), Daejeon, 34113, Rep. of Korea
[3] School of Electrical and Electronics Engineering, Chung-Ang University, Seoul, 06974, Rep. of Korea

*Author to whom any correspondence should be addressed.

**E-mail:** ysj@kriss.re.kr

## Abstract

In this study, we evaluate the measurement uncertainty of spectral domain interferometry with super-Gaussian windowing, which has significant potential for interferometer-based dimensional calibration services owing to its simple configuration and nanometer-level precision. A super-luminescent diode (SLD) with a broadband spectrum was used as the light source for high-precision length measurement, and a frequency-calibrated spectrometer captured the spectral interference signal. The length measurement results were compared with a laser displacement interferometer. To evaluate the measurement uncertainty, numerical simulations were performed by considering the intensity-noise level of the CCD in the spectrometer and the frequency-axis uncertainty of the spectrometer. Through a comprehensive uncertainty analysis supported by numerical simulations, the combined standard uncertainty of the measured length was evaluated as $\sqrt{(10\ \mathrm{nm})^2 + (9.7 \times 10^{-6} L)^2}$. The resulting uncertainty budget provides a metrological basis for extending SLD-based SDI with super-Gaussian windowing toward interferometric dimensional calibration tasks over the millimeter-scale range.

## 1. Introduction

Length is one of the fundamental physical quantities in the SI system of units, and the ability to measure it accurately is essential across science and technology [1-3]. Currently, the definition of the meter is practically realized by measuring the propagation time of light in vacuum or by interferometric length measurements using laser sources whose optical frequencies are traceable to the CIPM recommended values of standard frequencies [4,5], thereby supporting various dimensional calibration services [6,7]. However, laser displacement interferometry has a practical limitation because it determines distance by accumulating displacement from a starting point. Although various techniques can determine distance without accumulating displacement, conventional methods such as Time-of-Flight (ToF) [8], amplitude-modulated continuous wave (AMCW) [9], and frequency-modulated continuous wave (FMCW) [10] exhibit measurement uncertainties in the mm to μm range, making them unsuitable for calibration services. In the case of frequency-comb-based distance measurements [11], various methods such as dual-comb [12], multi-wavelength interferometry [13,14], pulse-to-pulse [15], and microcomb ranging [16] have been proposed with nanometric measurement precision; however, their system complexity currently makes them difficult to implement in practical calibration services. Spectral domain interferometry [17] is an interferometric technique capable of measuring length with nanometric precision without displacement accumulation. [18,19] Due to its simple configuration, it has been widely used across various industrial and scientific fields, including optical coherence tomography (OCT) imaging [20], optical-pulse dynamics measurements [21], long-distance measurement [22-26], and wafer thickness measurement [27-31].

In our previous work [19], we presented the effects of frequency and intensity noise on measurement precision. In this study, we evaluated the measurement uncertainty of spectral domain interferometry (SDI), considering the frequency and intensity uncertainties of the spectral interference signal. For high-precision length measurement, a broadband super-luminescent diode (SLD), super-Gaussian windowing and a high-speed spectrometer with a 40 kHz sampling rate were employed. The spectrometer was calibrated using a $C_2H_2$ laser-referenced wavelength meter to establish metrological traceability. To minimize artifacts in the length-domain data reconstructed via fast Fourier transform (FFT), a Super-Gaussian window was applied as an apodization function to the normalized interference spectrum [32-34]. To evaluate the repeatability, a fixed length was measured over 10 000 s, and the repeatability was found to be 1 nm in terms of Allan deviation. We experimentally confirmed that the measurement results of the SDI were consistent with those of the laser displacement interferometer within 16.5 nm (1 σ) over a range of 4 mm within the roll-off region. To evaluate the measurement uncertainty in SDI using an SLD and super-Gaussian windowing, a numerical simulation was conducted by considering intensity uncertainty of the CCD used in the spectrometer and the frequency uncertainty of the spectrometer. The measurement uncertainty was numerically evaluated to be $\sqrt{(10\ \mathrm{nm})^2+(9.7\times10^{-6}L)^2}$ ($k$=1). Together with the nanometer-level measurement precision, the uncertainty evaluation suggests that SDI using an SLD and super-Gaussian windowing could serve as a complementary technique for dimensional calibration services currently based on laser displacement interferometry or closely related interferometric length measurements.

## 2. Basic principle and experimental configuration, and data processing

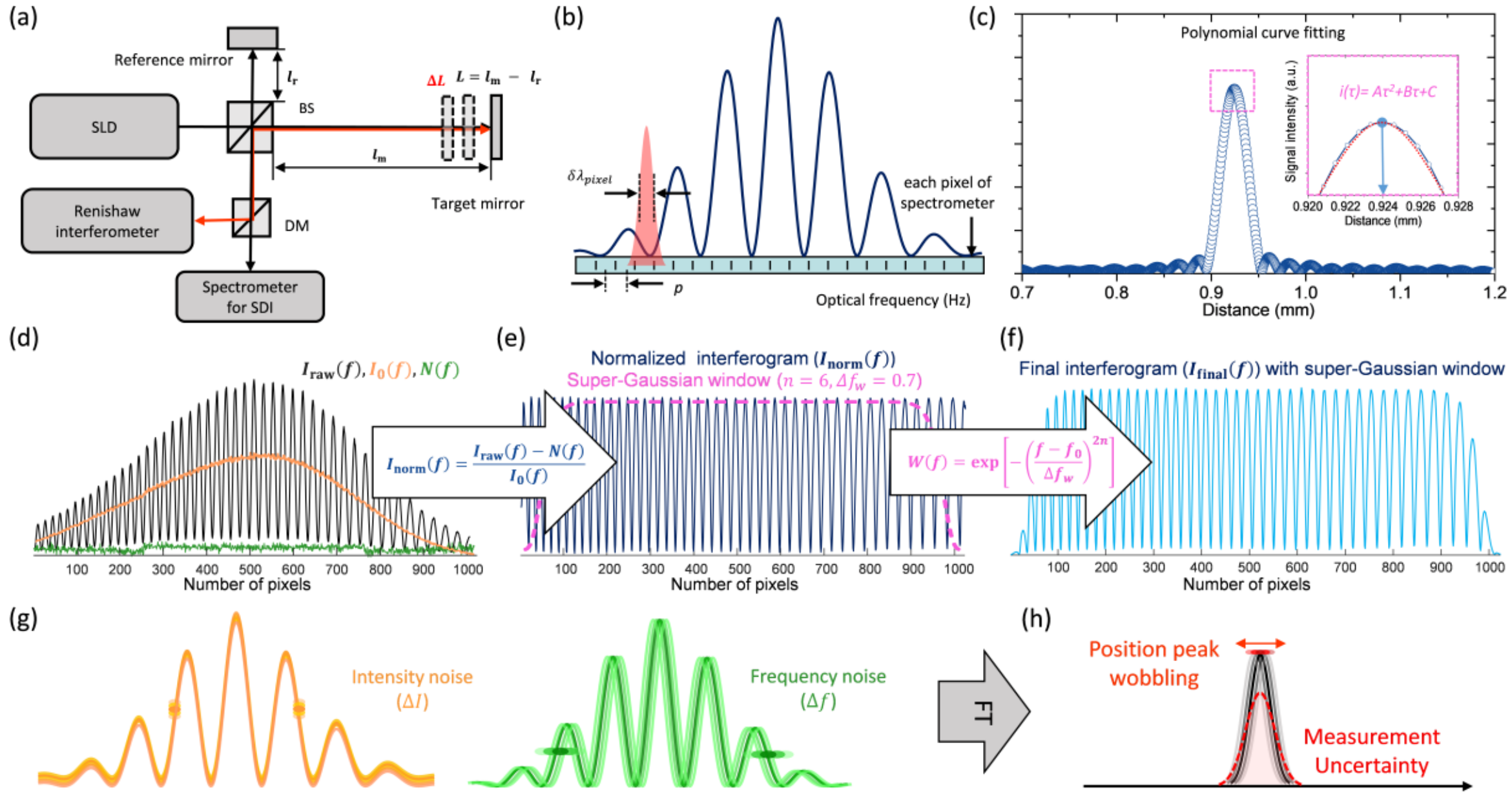

Fig. 1. Principle and signal-processing procedure of spectral domain interferometry (SDI) for length measurement. (a) Experimental configuration of the SDI and the reference laser displacement interferometer in a coaxial arrangement. (b) Spectral interference fringe sampled by the finite pixel width of the spectrometer, which constrains the practical measurable range. (c) Length-domain peak extraction after Fourier transformation and sub-pixel polynomial fitting. (d) Raw interferogram, spectral intensity

profile, and background-noise component used for normalization. (e) Normalized interferogram after source-spectrum normalization. (f) Super-Gaussian window ( $n = 10$, $\Delta f_w = 0.7$ ) applied to the normalized interferogram. (g) Fringe distortion induced by intensity and frequency variations. (h) Conversion of peak-position variation into measurement uncertainty.

The experimental configuration of the proposed spectral domain interferometer (SDI) is based on a Michelson-type interferometric structure, as illustrated in Fig. 1(a). To facilitate performance validation, a reference Renishaw laser displacement interferometer was used alongside the SDI for simultaneous length measurement. Both systems were aligned coaxially with the target's central axis. This alignment strategy effectively minimizes the cosine error by ensuring that the measurement beam is parallel to the displacement axis, while simultaneously mitigating the Abbe error by reducing the Abbe offset—the spatial separation between the measurement axis and the target's central axis. This alignment optimization ensures high measurement linearity and synchronization, facilitating a robust comparative assessment of system accuracy.

A super-luminescent diode (SLD) with a central wavelength of 1553.89 nm and a 3-dB bandwidth of $\Delta\lambda$ = 90 nm was employed as the broadband light source. The output beam is split into a reference arm and a measurement arm by a beam splitter. The measurement arm is equipped with a high-precision translation stage, allowing displacement comparison with a reference laser interferometer. The combined interference signal is captured by a high-speed spectrometer featuring a diffraction grating and a CCD array with 1024 pixels. To establish metrological traceability, the spectrometer’s wavelength-to-pixel mapping was precisely calibrated using a wavelength meter referenced to an acetylene ($C_2H_2$) stabilized laser. Further details regarding this configuration are described in Section 3.1.

The spectral interferogram captured by the spectrometer $I(f)$ is modulated by the optical path difference (OPD), $2nL$, between the two arms and can be expressed as:

$$I(f) = I_0(f)[1 + V(f)\cos\frac{4\pi nLf}{c}] \quad \text{(Eq.1)}$$

where $I_0(f)$ represents the spectrum of the SLD and $V(f)$ is the fringe visibility function. In an SDI system, the practical maximum measurable range, $L_{\max}$, is primarily constrained by the effective spectral resolution of the spectrometer, defined as the spectral width of a single pixel ($\delta f_{pixel}$) as shown in Fig. 1(b). Each pixel integrates the interference signal over its finite spectral width. The length at which fringe visibility significantly diminishes is given by [35]:

$$L_{max} = \frac{c}{4\delta f_{pixel}} \quad \text{(Eq.2)}$$

Our spectrometer covers a frequency range of 185.92 THz to 200.49 THz (1495.29 nm to 1612.49 nm) across 1024 pixels, corresponding to a spectral resolution ($\delta f_{pixel}$) of 14.2 GHz (0.11 nm). Based on Eq.(2), $L_{max}$ is theoretically determined to be 5.27 mm. However, because the fringe contrast decreases as the target length increases owing to the finite coherence length, and sidelobe artifacts also occur near $L_{max}$ from the conjugate plane in the FFT domain, the measurement range of our system was practically limited to 4 mm based on these two considerations.

The absolute length is evaluated by transforming the measured interferogram into the time (length, $L = c * \tau / n$) domain through the Fourier transform (FT) with $N_{\mathrm{FT}}$ zero-padding, as illustrated in Fig. 1(b). This numerical transformation dictates the digital length resolution ($\Delta z$), representing the spacing between adjacent discrete bins on the z-axis. Sub-pixel peak fitting enables nanometer-level precision beyond the constraints of the discrete digital bin size. As shown in Fig. 1(c), the interference peak in the Fourier domain is first filtered, and the data points adjacent to the maximum intensity are subsequently fitted to a quadratic polynomial, $i(\tau) = A\tau^2 + B\tau + C$. The exact peak center is extracted by finding the zero-crossing of the derivative, $\frac{di(\tau)}{d\tau} = 2A\tau + B = 0$, which improves peak localization and supports stable high-precision metrology.

For accurate signal analysis, the acquired interferogram is normalized in the frequency domain. First, the background noise is subtracted from the raw signal $I_{\mathrm{raw}}(f)$ , and the result is divided by the spectral envelope of the light source as shown in Fig. 1(d) and 1(e). The normalized interferogram $I_{\mathrm{norm}}(f)$ is defined as:

$$I_{\mathrm{norm}}(f) = \frac{I_{\mathrm{raw}}(f) - N(f)}{I_0(f)} \quad \text{(Eq.3)}$$

where $I_{\mathrm{raw}}(f)$ is the measured intensity, $N(f)$ is the dark noise offset, and $I_0(f)$ is the measured spectral intensity profile of the light source in the frequency domain. As illustrated in Fig. 1(f), to suppress sidelobe artifacts and enhance spectral linearity, a super-Gaussian windowing function $W(f)$ is applied for apodization:

$$W(f) = \exp\left[-\left(\frac{f - f_0}{\Delta f_w}\right)^{2n}\right] \quad \text{(Eq.4)}$$

where $f_0$ is the central frequency, $\Delta f_{\mathrm{w}}$ is the bandwidth parameter of the window, and $n$ is the order of the Super-Gaussian function. The final processed signal, $I_{\mathrm{final}}(f) = I_{\mathrm{norm}}(f) \cdot W(f)$, is then transformed into the length domain via a Fourier transform for sub-pixel peak fitting. For the SLD source and spectrometer used in this study, the Super-Gaussian window parameters were set to $n = 10$ and $\Delta f_w = 0.7$, which gave the highest length-measurement accuracy and were therefore used in the subsequent data processing.

$$L = \frac{c\tau(I, f)}{2n_{\mathrm{air}}} \quad \text{(Eq.5)}$$

The final length expression is therefore given by Eq. (5). The peak position in the time-delay domain, $\tau(I, f)$, is extracted from the Fourier transform of the final processed interferogram. This expression explicitly indicates that the estimated delay depends on both the measured intensity spectrum $I$ and the calibrated optical-frequency axis $f$ used for resampling and FFT processing. As illustrated in Fig. 1(g) and 1(h), intensity noise distorts the fringe envelope and affects the fitted peak position, whereas optical-frequency noise and frequency-axis sampling error change the conversion from the spectral interferogram to the delay-domain peak. These effects therefore propagate directly to the length uncertainty.

## 3. Experimental results

### 3.1. Frequency calibration of the spectrometer

For metrological traceability, the wavelength-to-pixel mapping was calibrated using a tunable laser monitored by a wavelength meter that had been pre-calibrated against an acetylene ($C_2H_2$) stabilized laser. The wavelength meter used as the reference instrument for this frequency calibration had a relative standard uncertainty of $2.2\times10^{-7}$ ($k$ = 1). The 1024-pixel mapping was established by sweeping the laser from 1498 nm to 1612 nm in 2 nm increments, with the wavelength meter readings serving as the reference for mapping frequency to each pixel.

To ensure the reliability of the calibration model and prevent overfitting, the dataset was partitioned into calibration and validation sets for cross-validation arrangement (Fig. 2(a)). A comparative analysis of various polynomial orders led to the selection of a fourth-order polynomial fit, which exhibited the lowest residual error. Evaluation of the model using validation data (Fig. 2(b)) revealed a residual error of less than 0.01 nm (approx. 1.2 GHz). The calculated standard deviation was 6 pm (0.7 GHz), which corresponds to a relative frequency uncertainty of $3.8 \times 10^{-6}$. Note that the frequency uncertainty of the spectrometer can be further improved by frequency-comb-based calibration [36].

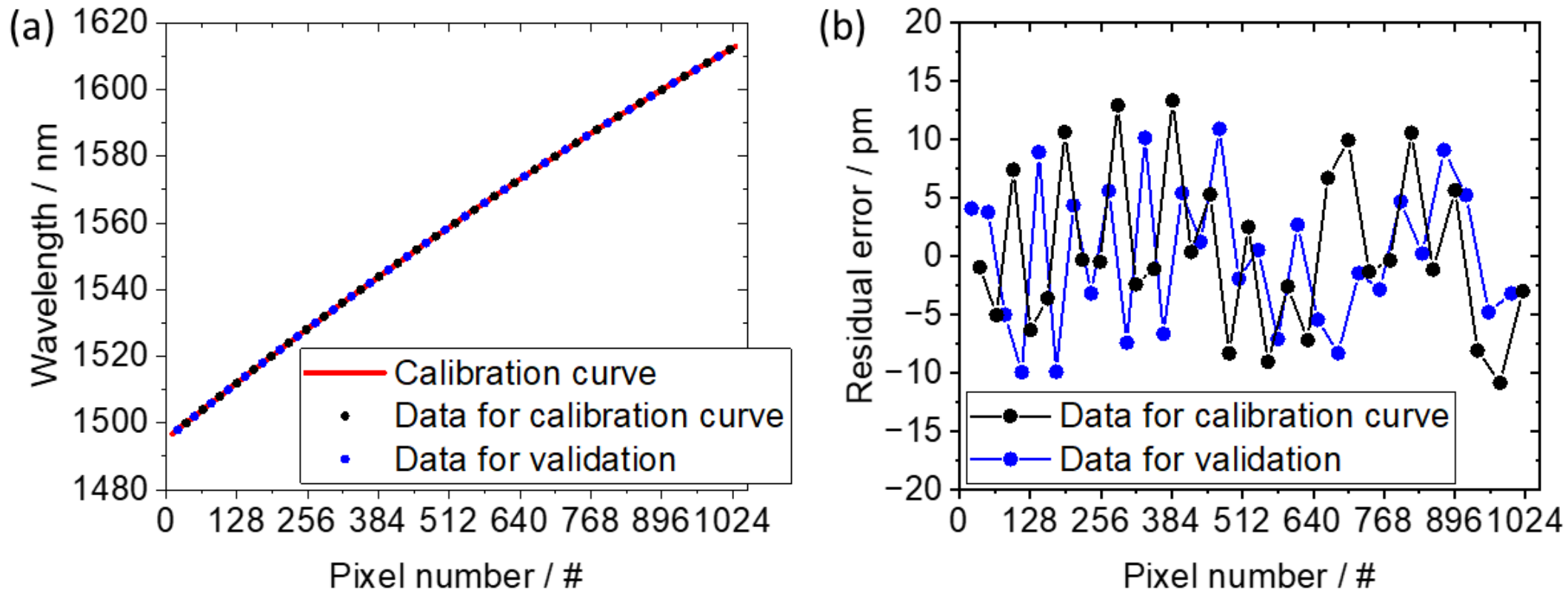

Fig. 2. Spectrometer frequency calibration results using a tunable laser and a wavelength meter: (a) Cross-validation arrangement of calibration and validation data showing wavelength measurements per pixel, (b) Residual error distribution of calibration and validation data after applying the fourth-order polynomial fit.

### 3.2. Repeatability of length measurement

To evaluate measurement repeatability, length measurements were performed at a fixed length of approximately 750 μm. The optical path length (OPL) of the system was minimized and the optical path was enclosed to suppress temperature-dependent optical-path drift. First, to determine the precision limit of the system, 40 000 data points were acquired at a sampling rate of 40 kHz over a duration of 1 s. For this short-term measurement, the standard deviation of the measured length was found to be 1.5 nm as shown in Fig. 3(a). Second, Fig. 3(b) shows the characteristics of the long-term drift, measurements were conducted at a 1 Hz update rate for 10 000 s. For this long-term measurement, the standard deviation of the measured length was 4.2 nm, with no notable drift observed.

To quantify the measurement repeatability, the Allan deviation was calculated as illustrated in Fig. 3(c). The Allan deviation of the measured length was approximately 1 nm at an averaging time of 25 μs, which gradually improved to 0.3 nm at an averaging time of 1 ms. From 25 μs to 1 ms, the experimental results were in excellent agreement with the white-noise-limited Allan deviation, represented by 5.0 pm$\times\tau^{-1/2}$. For averaging times longer than 1 ms, the Allan deviation increased gradually due to random walk noise induced by thermal expansion in the laboratory environment, ultimately reaching about 1 nm for averaging times longer than 1 s. [19,36] Consequently, we conservatively estimate our measurement repeatability as 1 nm under standard laboratory conditions. Although the system's intrinsic repeatability is estimated to be well below 1 nm, we believe this specification provides a more realistic representation of its performance in practical environments.

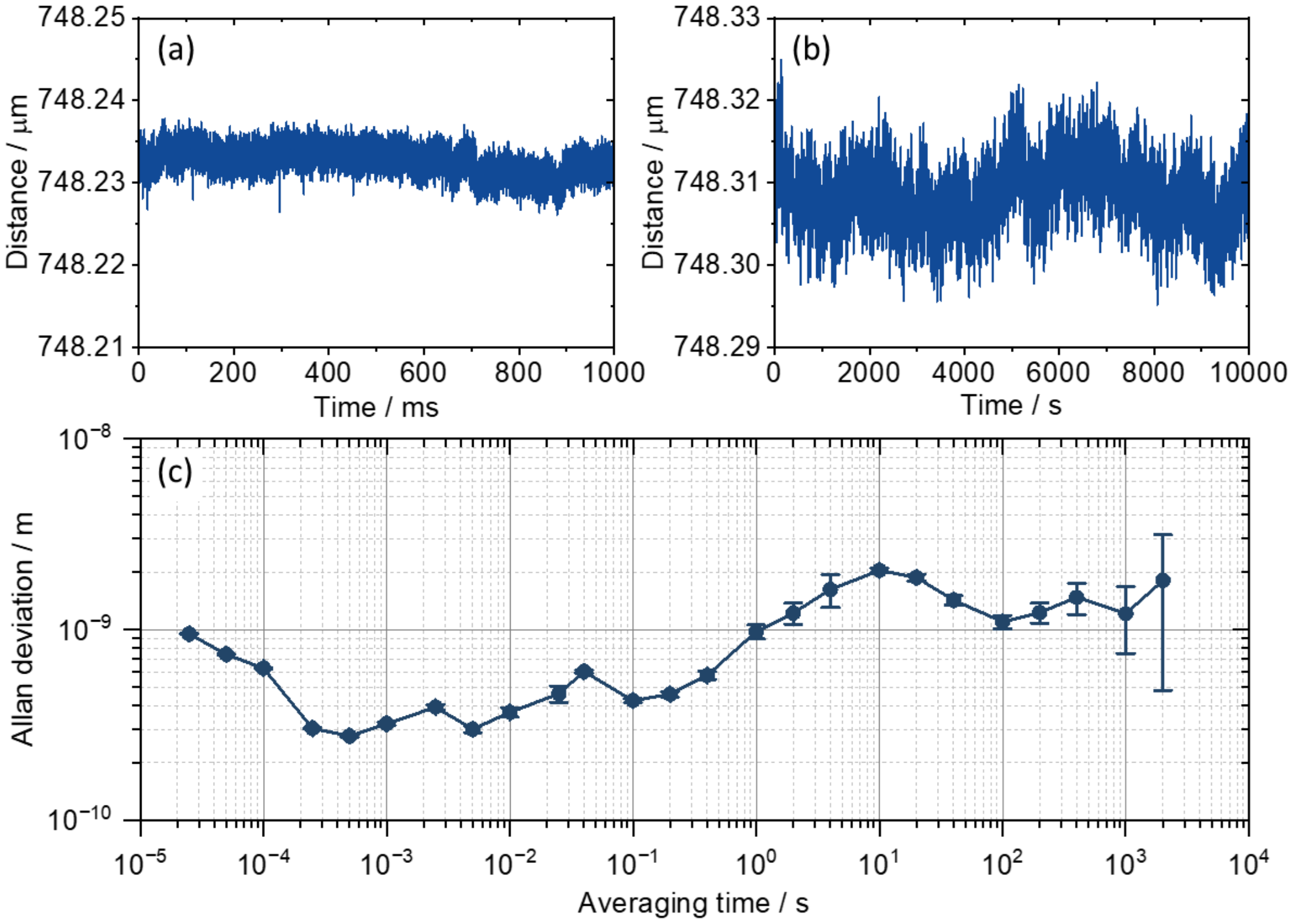

Fig. 3. Repeatability evaluation. (a) Length measurement with an update rate of 40 kHz over 1 s. (b) Length measurement with an update rate of 1 Hz over 10 000 s. (c) Allan deviation from 25 µs to 2 000 s to evaluate the measurement repeatability.

### 3.3. Measurement linearity comparison with laser displacement interferometer

To evaluate the measurement linearity, a comparison experiment was performed with a laser displacement interferometer (Renishaw, model RLE20) with a measurement linearity of 5 nm. A target mirror was translated over 4 mm with 100 µm steps, and a trigger signal was used to synchronize measurements between the laser displacement interferometer and the spectral domain interferometer, thereby minimizing the influence of drift during the measurement. As shown in Fig. 4(a), we experimentally confirmed that the measurement results of the spectral domain interferometer were consistent with those of the laser displacement interferometer within 16.5 nm (1 σ) over a range of 4 mm without notable Abbe or cosine error. To validate these experimental results, a numerical simulation was performed. We modelled these measurement uncertainty components by assuming that they arise from intensity and frequency uncertainties of the measured spectral interference signal, which prevent the acquisition of an ideal spectral interference signal [19]. The previously obtained frequency-axis error ($\Delta f/f$) and an intensity noise ($\Delta I/I$) level of 1.6 % of the full-scale signal, specified by the manufacturer, were introduced into an ideal interference spectrum to generate the Monte Carlo distribution shown by the gray dots.

The individual contributions of the intensity noise and frequency-axis error terms were then evaluated separately, and their expanded uncertainties ($k$ = 2) are represented by the blue and green dashed lines, respectively. The contribution of the intensity uncertainty to the measured length (blue dashed line) was evaluated to be approximately 10 nm ($k$ = 1), independent of the measured length, whereas the contribution of the frequency-axis error to the measured length (green dashed line) can be expressed by $9.7 \times 10^{-6} L$ ($k$ = 1), showing a linear dependence on the measured length. This behavior is illustrated by the Monte Carlo results in Fig. 4(b) and Fig. 4(c): the intensity noise contribution is dominant at 0.23 mm, while the frequency-axis error contribution is dominant at 4 mm. The red dashed line represents the combined standard uncertainty from the two contributions. The agreement between the experimental results and the numerically estimated uncertainty indicates that the simulation adequately represents the system and that the evaluated uncertainty is reasonable.

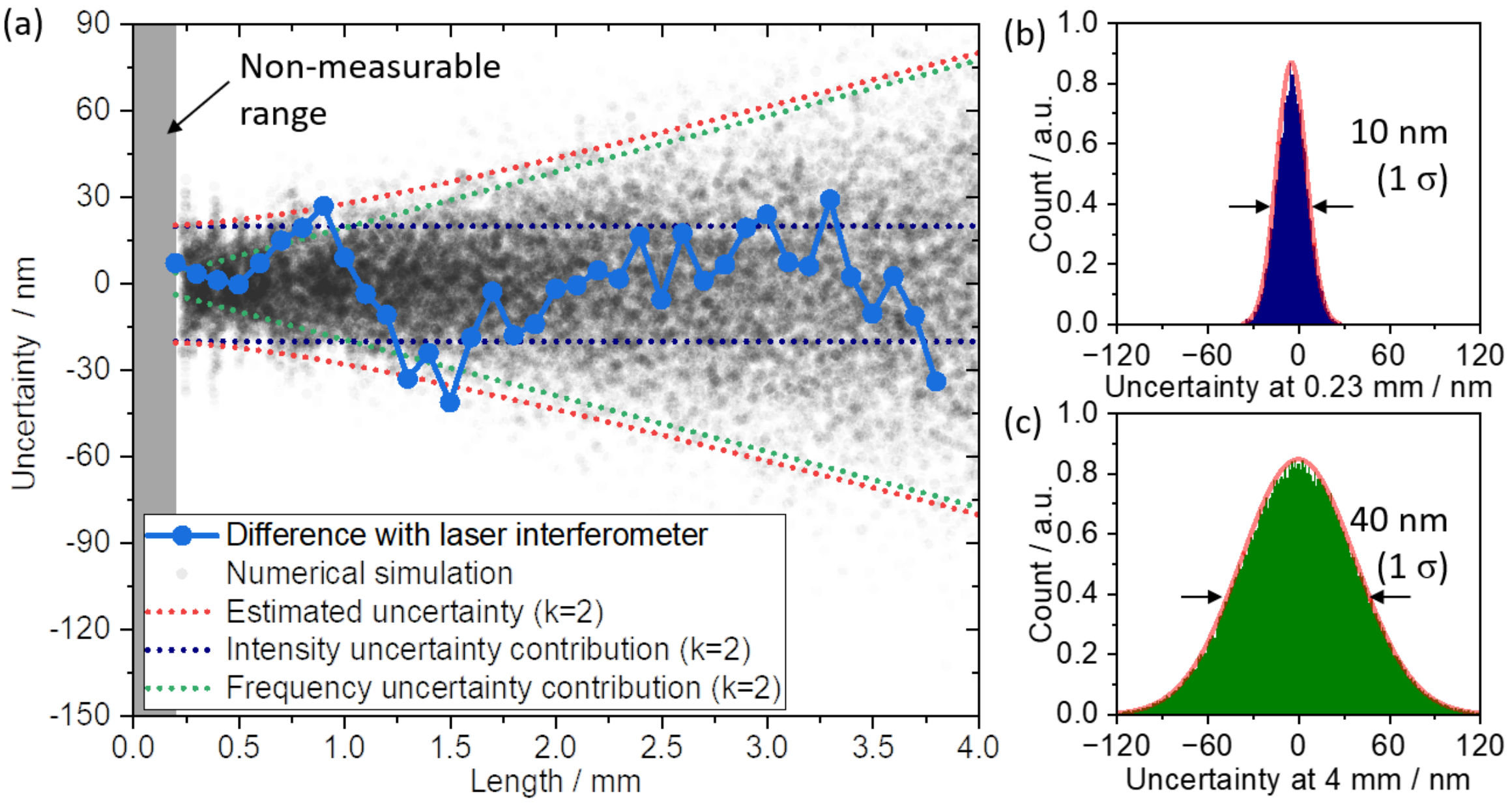

Fig. 4. Linearity comparison and numerical simulation-based uncertainty analysis. (a) Experimental results and numerical simulation results. (b) Numerical simulation of uncertainty evaluation at 0.23 mm. (c) Numerical simulation of uncertainty evaluation at 4 mm.

## 4. Evaluation of measurement uncertainty

To evaluate the measurement uncertainty, the measurement function was formulated according to the Guide to the Expression of Uncertainty in Measurement (GUM). As described in Eq. (5), the length in the proposed spectral domain interferometer is obtained from the round-trip delay extracted from the processed spectral interferogram. Because the time delay, $\tau(I, f)$, is obtained through normalization, apodization, FFT, and sub-pixel peak fitting, its dependence on the input data cannot be written as a simple closed-form expression. Therefore, for uncertainty propagation, the measurand is written as an implicit measurement function:

$$L = f(I, f, n_{\mathrm{air}}) \quad \text{(Eq.6)}$$

Under the assumption that the intensity, frequency and air-refractive-index terms are evaluated independently, the combined standard uncertainty can be expressed as

$$u_{\mathrm{c}}^2(L) = c_{\mathrm{I}}^2 u^2(I) + c_{\mathrm{f}}^2 u^2(f) + c_{\mathrm{n}}^2 u^2(n_{\mathrm{air}}) + u_{\mathrm{proc}}^2 + u_{\mathrm{rep}}^2 \quad \text{(Eq.7)}$$

In Eq. (7), $c_{\mathrm{I}}$ and $c_{\mathrm{f}}$ are sensitivity coefficients for the intensity and frequency terms, respectively. Since these two terms pass through the full signal-processing chain, they were evaluated numerically rather than by closed-form differentiation. The air-refractive-index coefficient is obtained analytically from Eq. (5), whereas the intensity and frequency contributions are obtained from Monte Carlo simulations as

$$c_{\mathrm{n}} = \frac{\partial L}{\partial n_{\mathrm{air}}} = -\frac{L}{n_{\mathrm{air}}}$$
$$c_{\mathrm{I}} u(I) \equiv u_I(L) = s[L_k^{(I)}] \quad \text{(Eq.8)}$$
$$c_{\mathrm{f}} u(f) \equiv u_f(L) = s[L_k^{(f)}]$$

To make the numerical evaluation explicit, the values in Table 1 were substituted into Eq. (7) component by component. The spectrometer terms used the assigned intensity noise $\Delta I/I = 1.6\ \%$ and the calibrated frequency-axis uncertainty $\Delta f/f = 3.8 \times 10^{-6}$. Since these two inputs were treated as independent, they were simulated separately with 20 000 Monte Carlo trials.

Each set of trials, either $I$ or $f$ was perturbed and then applied the same normalization, apodization, FFT, and peak-fitting process used for $L$. The standard deviations of the recalculated $L$ values give $c_\mathrm{I} = 10\ \mathrm{nm}/1.6\ \%$ and $c_\mathrm{f} = 2.6 \times L$; hence $u_I(L) = 10$ nm and $u_f(L) = (2.6 \times L) \cdot (3.8 \times 10^{-6}) = 9.7 \times 10^{-6} L$. This frequency term corresponds to the calibrated spectrometer frequency-axis uncertainty and may be reduced by frequency-comb-based calibration.

$$u_{\mathrm{spec}}(L) = \sqrt{u_I(L) + u_f(L)} = \sqrt{(10\ \mathrm{nm})^2 + (9.7 \times 10^{-6} L)^2}$$

During the uncertainty evaluation, the measurement environment was passively stabilized to reduce variations in the refractive index of air. Air temperature and pressure were monitored using environmental sensors with standard uncertainties of 41 mK and 29 Pa, respectively, and these values were used in the Ciddor-equation-based evaluation of the residual air-refractive-index uncertainty [37-39]. The resulting contributions were $1.0\times10^{-8}\times L$ from the Ciddor equation, $3.8\times10^{-8}\times L$ from air temperature $(u(T) = 41\ \mathrm{mK})$, $7.8\times10^{-8}\times L$ from air pressure $(u(P) = 29\ \mathrm{Pa})$, $4.8\times10^{-9}\times\mathrm{L}$ from humidity $(u(RH) = 0.6\ \%\mathrm{RH})$, and $6.0\times10^{-9}\times L$ from $CO_2$ concentration $(u(CO_2) = 40\ \mathrm{ppm})$. Their root-sum-square gives $u(n_{air}) = 8.7 \times 10^{-8}$ and $u_{n_\mathrm{air}}(L) = 8.7 \times 10^{-8} L$, where $c_\mathrm{n} = L\ /\ n_\mathrm{air} \approx L$ because $n_\mathrm{air} \approx 1$.

Finally, the 1.3 nm data-processing term $(u_{proc})$, the spectrometer term $u_\mathrm{spec}(L)$, the 1.0 nm repeatability term $(u_{rep})$ [19,36], and the air-refractive-index term $(u_{n_\mathrm{air}}(L))$ were combined in quadrature:

$$u_c(L) = \sqrt{(10\ \mathrm{nm})^2 + (9.7 \times 10^{-6} L)^2}$$

For L = 4 mm, the frequency contribution is 38.8 nm and the air-refractive-index contribution is 0.35 nm. Substituting these values into the combined budget gives $u_\mathrm{c}(L) = 40$ nm, as listed in Table 1.

| Category | Sources | Standard uncertainty $u(x_i)$ | Sensitivity coefficient ($c_{x_i}$) | Contribution $u_i(L)$ | Type |
|---|---|---|---|---|---|
| Data processing | Normalization, Windowing, FFT process | 1.3 nm | 1 | 1.3 nm | B |
| Spectrometer | | $\sqrt{(10\ nm)^2 + (9.7 \times 10^{-6} L)^2}$ | 1 | $\sqrt{(10\ nm)^2 + (9.7 \times 10^{-6} L)^2}$ | B |
| | Intensity | $\Delta I/I =$ 1.6 % | 10 nm / 1.6 % | 10 nm | B |
| | Frequency | $\Delta f/f$= $3.8 \times 10^{-6}$ | $2.6 \cdot L$ | $9.7 \times 10^{-6} L$ | B |
| Repeatability | | 1.0 nm | 1 | 1.0 nm | A |
| Refractive index of air | | $8.7 \times 10^{-8}$ | L | $8.7 \times 10^{-8} L$ | B |
| | Ciddor equation | $1.0 \times 10^{-8}$ | L | $1.0 \times 10^{-8} L$ | B |
| | Air temperature | 41 mK | $9.3 \times 10^{-7} L \times K^{-1}$ | $3.8 \times 10^{-8} L$ | B |
| | Air pressure | 29 Pa | $2.7 \times 10^{-9} L \times Pa^{-1}$ | $7.8 \times 10^{-8} L$ | B |
| | Humidity | 0.6 %RH | $8.0 \times 10^{-9} L \times \%RH^{-1}$ | $4.8 \times 10^{-9} L$ | B |
| | $CO_2$ | 40 ppm | $1.5 \times 10^{-10} L \times ppm^{-1}$ | $6.0 \times 10^{-9} L$ | B |
| Combined standard uncertainty $u_c$ | | | | $\sqrt{(10\ \mathrm{nm})^2 + (9.7 \times 10^{-6} L)^2}$ | |
| Expanded uncertainty $U$ | | | | $\sqrt{(20\ \mathrm{nm})^2 + (1.9 \times 10^{-5} L)^2}$ ($k$=2) | |

## 5. Conclusion

In this study, we evaluated an uncertainty for length measurement by spectral domain interferometry using a broadband SLD source and a frequency-calibrated spectrometer with super-Gaussian windowing. By analytically for formulating the uncertainty relations from the measurement function, we performed a rigorous uncertainty evaluation based on experimental results in accordance with the GUM. The intensity and frequency contributions were obtained numerically by Monte Carlo simulations using the same normalization, super-Gaussian windowing, FFT, and peak-fitting procedure as the measurement, while the air-refractive-index contribution was evaluated from the Ciddor equation and environmental sensor uncertainties. Combining the data-processing, repeatability, spectrometer, and air-refractive-index terms yielded an expanded standard uncertainty of 80 nm ($k$ = 2) at L = 4 mm, consistent with the comparison against a traceable laser displacement interferometer and the numerical simulation results.

These results show that spectral domain interferometry, when supported by traceable spectrometer calibration and an explicit uncertainty budget, can provide nanometer-level uncertainty over the millimeter-scale measurement range. Furthermore, the systematic approach to uncertainty evaluation presented in this work is expected to provide a practical guideline for researchers working with spectral interferometry. In particular, the method could complement calibration tasks that are presently traceable through laser displacement interferometry or closely related interferometric length measurements, such as standard scales, gauge blocks, high-precision displacement transducer such as LVDT and capacitive sensor, wafer thickness and the roundness or form standard specimens. Further reduction of the spectrometer frequency-axis uncertainty, for example by frequency-comb-based calibration, combined with application-specific uncertainty evaluation, is expected to improve the traceability and practical applicability of this approach for a broader range of dimensional calibration services.

## Acknowledgements

This research was supported by the Development of World-Leading Technology in Physical Measurement Standards funded by Korea Research Institute of Standards and Science (KRISS – 2026 – GP2026 - 0001) and the National Research Foundation of Korea (NRF) grant funded by the Korea government (MSIT) (No. RS-2026-25483063, RS-2025-16043000).